\documentclass[conference]{IEEEtran}
\IEEEoverridecommandlockouts
\usepackage{cite}
\usepackage{amsmath,amssymb,amsfonts}
\usepackage{algorithmic}
\usepackage{graphicx}
\usepackage{textcomp}
\usepackage{xcolor}

\usepackage{subcaption}
\usepackage{xspace}
\usepackage{soul}

\usepackage{url}
\usepackage[hidelinks]{hyperref}
\usepackage[capitalise]{cleveref}

\usepackage{algorithm}
\usepackage{algorithmic}
\usepackage{wrapfig}
\usepackage{multirow}

\usepackage{lipsum}
\usepackage{makecell}
\usepackage[export]{adjustbox}

\usepackage{tabularx}
\usepackage{array}

\usepackage[table,xcdraw]{xcolor}

\newcolumntype{Y}{>{\raggedright\arraybackslash}X}

\usepackage{fancyhdr}

\newif\ifdraft 
\draftfalse

\newcommand{\name}{EnergAIzer\xspace}

\def\BibTeX{{\rm B\kern-.05em{\sc i\kern-.025em b}\kern-.08em
    T\kern-.1667em\lower.7ex\hbox{E}\kern-.125emX}}
\begin{document}


\title{\name: Fast and Accurate GPU Power Estimation Framework for AI Workloads \\
\thanks{This work was supported by the Massachusetts Institute of Technology-International Business Machines
(MIT-IBM) Watson AI Laboratory.}
}

\author{

\IEEEauthorblockN{Kyungmi Lee$^1$, Zhiye Song$^1$, Eun Kyung Lee$^2$, Xin Zhang$^2$, Tamar Eilam$^2$, Anantha P. Chandrakasan$^1$}
\IEEEauthorblockA{$^1$ \textit{Massachusetts Institute of Technology, Cambridge, MA, USA} \\
$^2$ \textit{IBM Research, Yorktown Heights, NY, USA} \\
\texttt{\{kyungmi,zhiye\}@mit.edu}, \texttt{\{eunkyung.lee,xzhang,eilamt\}@us.ibm.com}
}

}

\sloppy
\maketitle
\thispagestyle{empty}
\begin{abstract}
As AI workloads drive increases in datacenter power consumption, accurate GPU power estimation is critical for proactive power management. 
However, existing power models face a scalability bottleneck not in the modeling techniques themselves, but in obtaining the hardware utilization inputs they require. 
Conventional approaches rely on either costly simulation or hardware profiling, which makes them impractical when rapid predictions are required. 

This work presents \name, which addresses this scalability bottleneck by developing a lightweight solution to predict utilization inputs, reducing the estimation walltime from hours to seconds. 
Our key insight is that kernels in AI workloads commonly employ optimizations that create structured patterns, which analytically determine memory traffic and execution timeline. 
We construct a performance model using these patterns as an analytical scaffold for empirical data fitting, which also naturally exposes module-level utilization. 
This predicted utilization is then fed into our power model to estimate dynamic power consumption. 

\name achieves 8\% power errors on NVIDIA Ampere GPUs, competitive with traditional power models with elaborate cycle-level simulation or hardware profiling. 
We demonstrate \name's exploration capabilities for frequency scaling and architectural configurations, including forecasting the power of NVIDIA H100 with just 7\% error. 
In summary, \name provides fast and accurate power prediction for AI workloads, paving the way for power-aware design explorations.

\end{abstract}


\section{Introduction} \label{sec:intro}

Artificial intelligence workloads are driving significant increases in power and energy demand across global datacenters.
Datacenters are projected to consume up to 12\% of US electricity by 2028~\cite{lbnldatacenter}, with AI workloads accounting for 27\% of datacenter power by 2030~\cite{mitaidatacenter}. 
Graphics processing units (GPUs), the primary accelerators for AI workloads, are the dominant source of the power consumption~\cite{gpupowerfraction}, with thermal design power (TDP) reaching 700W and 1200W in recent NVIDIA H100~\cite{h100} and GB200~\cite{GB200}, respectively.

In light of these growing energy challenges, rapid estimation of GPU power and energy consumption for AI workloads becomes critical. 
For instance, fast predictions allow datacenter operators to optimize resource allocation across multiple GPU configurations~\cite{yu2023,dynamollm,polca2024,kepler2023,energyawaretiling} and DVFS settings~\cite{zeus2023,perseus2024} without running every workload. 
Similarly, AI developers can forecast their workloads' energy consumption on new GPU configurations, enabling energy-aware design before new hardware becomes widely accessible. 
 
Despite extensive research on GPU power modeling, rapid power prediction for AI workloads remains challenging. 
Researchers have developed sophisticated techniques, including microarchitectural analysis~\cite{gpuwattch2013,accelwattch2021,gpujoule2019,ipp2010}, DVFS-aware prediction~\cite{guerreiro2018}, and machine learning approaches~\cite{wu2015,ali2023,zhang2024}. 
However, \emph{the limitation lies not in the modeling techniques themselves, but in how these models obtain their inputs}. 
Power models require hardware utilization information as input, indicating how intensively GPU modules (e.g., DRAM, Tensor Cores) are used, since dynamic power consumption is proportional to the module activity~\cite{cmospower}. 

Existing approaches obtain this information through two methods.
First, instruction-level simulators~\cite{accelsim2020,cudnngpgpusim2019,cutalssgpgpusim2019,pka2021,chung2024allegro} emulate GPU execution cycle-by-cycle to derive module utilization.
However, this detailed simulation takes several hours even for moderate-sized workloads~\cite{pka2021,li2023micro}. 
Alternatively, runtime profiling executes the workload on a GPU to collect performance counters, which incurs substantial profiling overhead and requires a readily available GPU. 
These methods create a scalability bottleneck, making existing power models impractical for resource optimization and design space exploration.

In this work, we address this scalability bottleneck through a unified performance and power modeling framework. 
Specifically, we develop \emph{a lightweight performance model that directly provides hardware utilization information to the power model} without costly simulation or profiling. 

\vspace{0.3em}
\noindent\textbf{Challenge~}
Recent work has shown that lightweight performance models can achieve scalable and accurate latency predictions for AI workloads~\cite{neusight2025,li2023micro,habitat2021}.
These models abstract workloads in a coarse-grained manner, representing each kernel by high-level features like total FLOPs and memory footprint, instead of fine-grained instructions. 
They then use data-driven techniques to predict aggregate performance metrics like latency and throughput from these features. 
While promising for latency predictions, they face a critical limitation for power modeling. 

Two workloads can have similar latency but significantly different power consumption due to underlying utilization patterns, which these lightweight performance models cannot capture. 
For example, the memory accesses in GPUs have different energy costs, with DRAM consuming significantly more energy than L2 cache~\cite{horowitz}. 
However, prior lightweight models abstract the memory traffic as a total footprint, losing distinctions between L2 cache and DRAM for accurate power modeling. 

Similarly, load imbalance across streaming multiprocessors (SMs) implies that latency depends on the slowest unit, whereas power consumption depends on the average utilization across all units. 
Yet, prior models that abstract workloads as FLOPs and memory footprint per threadblock miss these load distribution effects. 
Therefore, our challenge is to balance two competing requirements when designing a lightweight model: maintaining \emph{coarse-grained abstraction for scalability} while capturing \emph{the utilization patterns necessary for accurate power modeling}.

\begin{figure}[t]
    \centering
    \includegraphics[width=\linewidth]{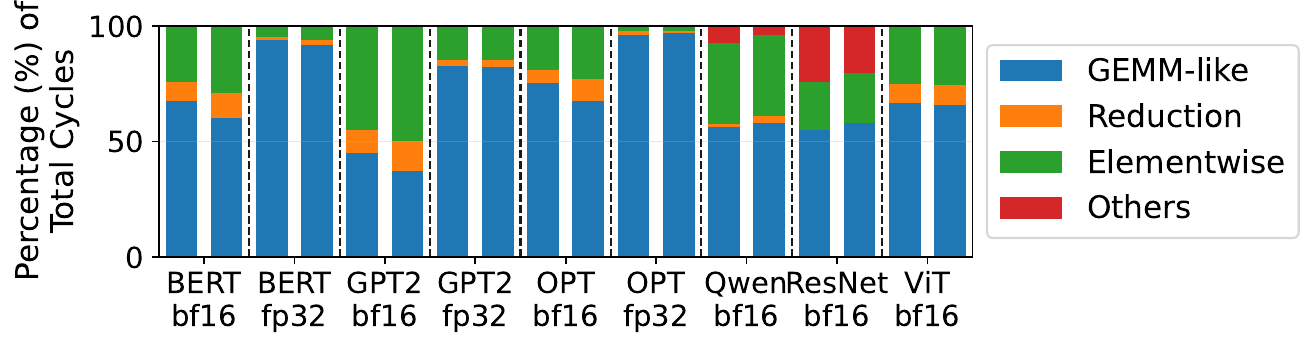}
    \vspace{-2mm}
    \caption{Latency breakdown of AI workloads according to the three kernel types: GEMM-like, Reduction, and Elementwise. Input shapes are omitted from the label.}
    \vspace{-4mm}
    \label{fig:background-kernel-breakdown}
\end{figure}

\vspace{0.3em}
\noindent\textbf{This Work~}
We present \name, a fast GPU power estimation framework for AI workloads. 
Our key insight is that AI workloads are dominated by kernels employing well-established software optimizations that create structured, analyzable patterns in hardware utilization. 
These kernels include generalized matrix multiplications (GEMMs), nonlinear reduction functions (e.g., softmax), and simple elementwise operations (e.g., activation functions), which together account for 90-99\% of execution time in several language and vision models (\cref{fig:background-kernel-breakdown}). 
The optimization choices of these kernels, including tiling across the execution hierarchy, threadblock scheduling, and pipelining, determine memory traffic, load distribution, and latency hiding. 
Thus, we leverage this structured pattern by abstracting each kernel through its optimization choice, which is coarse-grained yet captures architecturally relevant information for power modeling. 

Our performance model uses this abstraction as an analytical scaffold on which we apply data-driven refinement to account for GPU non-idealities. 
The analytical scaffold models execution as a timeline of coarse-grained actions (tiles and pipeline stages), naturally exposing each module's relative utilization.
For power modeling, we feed the predicted utilization as activity factors ($\alpha$) in the standard dynamic power equation (i.e., $\alpha CV^2f$), considering voltage-frequency scaling. 
\name achieves 8.8-11.0\% performance and 8.0-8.2\% power estimation errors on average across diverse language and vision workloads for NVIDIA Ampere GPUs. 

Finally, because our approach models software optimization patterns rather than their microarchitectural implementation details, it generalizes across recent GPU generations.
These generations differ primarily in configurations (number of SMs, memory bandwidth, Tensor Core throughput) while maintaining consistent execution models.
We demonstrate that \name forecasts the power consumption of NVIDIA Hopper and Lovelace GPUs with 7\% and 13\% error, respectively.

\section{Background} \label{sec:background}
\subsection{GPU Architecture} \label{sec:background-gpu}

GPU architecture is designed for parallelism, comprising multiple streaming multiprocessors (SMs), each equipped with configurable memory that can function as either shared memory or L1 cache. 
Each SM has four subpartitions (SMSPs), each including register files, Tensor Cores, CUDA Cores, and Special Function Units (SFUs). 
All SMs share global memory, which includes L2 cache and DRAM, serving as the main memory for the entire GPU. 

The software programs follow a hierarchical structure that maps to the GPU architecture~\cite{cuda}. 
Threads are the basic unit of computation, which are organized hierarchically: 32 threads form a warp, multiple warps are grouped into a threadblock, and multiple threadblocks complete the workload (grid). 
Warps execute in a SIMT (single instruction, multiple thread) fashion on SMSPs.
Multiple threadblocks run concurrently on each SM, limited by available hardware resources.
Threads within a threadblock share the SM's shared memory, enabling efficient data reuse and synchronization. 

\vspace{-0.4em}
\subsection{Power Modeling} \label{sec:background-power}

The power consumption of GPUs consists of dynamic power from transistor switching activity, static power from leakage currents~\cite{cmospower}, and constant power from board components~\cite{accelwattch2021}. 
Dynamic power is expressed as $\alpha CV^2f$, where $\alpha$ represents the activity factor, which is typically set as the module-level utilization (e.g., fraction of time a module is active)~\cite{accelwattch2021,gpujoule2019,guerreiro2018,ipp2010}.
$C$ is the hardware-specific parameter, and $V$ and $f$ are operating voltage and frequency.
Static power, along with constant board-level power consumption, forms an idle power that is largely independent of the workload and can be measured directly.
Therefore, we focus on accurately estimating the dynamic power in this work, which requires the utilization information $\alpha$ as the input. 

\section{Motivation and Objective} \label{sec:motivation} 

\subsection{Objective}

Our objective is to develop a rapid GPU power estimation framework for AI workloads.
AI workloads intensify power demands, making power a first-class design consideration that requires exploring numerous design points, such as GPU configurations, frequency scaling, and resource allocations. 

\subsection{Limitation of Existing Solutions}
\subsubsection{Scalability Challenge}\label{sec:motivation-scalability}
There has been extensive research on developing accurate GPU power models~\cite{gpuwattch2013,accelwattch2021,gpujoule2019,ipp2010,guerreiro2018,wu2015}, but obtaining the utilization inputs these models require remains a significant bottleneck. 
Conventionally, these inputs are derived through instruction-level performance simulators that emulate GPU behavior at the cycle level~\cite{pka2021,chung2024allegro,cudnngpgpusim2019,cutalssgpgpusim2019,accelsim2020}, or through profiling that executes workloads on physical GPUs to collect hardware counters. 

\begin{wraptable}{r}{0.5\linewidth}
\vspace{-4mm}
\captionof{table}{Walltime for Obtaining Utilization Inputs}
\footnotesize
\label{tab:sim-ncu}
\vspace{-2mm}
\begin{tabular}[t]{|c|c|c|}
\hline
\textbf{Model} & \textbf{Sim.}~\cite{pka2021} & \textbf{NCU} \\
\hline\hline
ResNet50 & 1.5 hr & 8.5 min \\\hline
BERT & 0.4 hr & 12.7 min \\\hline
Qwen2 & N/A & 35 min \\
\hline
\end{tabular}
\vspace{-4mm}
\end{wraptable}
Both approaches become scalability bottlenecks to power modeling. 
As shown in \cref{tab:sim-ncu}, an instruction-level simulation takes $0.4-1.5$ hours, while profiling using NVIDIA NSight Compute (NCU)~\cite{ncu} requires $8.5-35$ minutes per workload. 
Although these runtimes may be acceptable for detailed microarchitectural studies, they become prohibitive when evaluating multiple design points.

\subsubsection{Accuracy Challenge}\label{sec:motivation-accuracy}

To achieve scalability, one might consider recent lightweight performance models~\cite{neusight2025,li2023micro,habitat2021} that provide rapid latency estimation for AI workloads. 
They train machine learning models (e.g., neural networks~\cite{neusight2025,habitat2021}, linear regression~\cite{li2023micro}) to predict aggregate metrics like latency or throughput from high-level workload features, such as FLOPs and memory footprint. 

\begin{figure}[t]
    \centering
    \includegraphics[width=\linewidth]{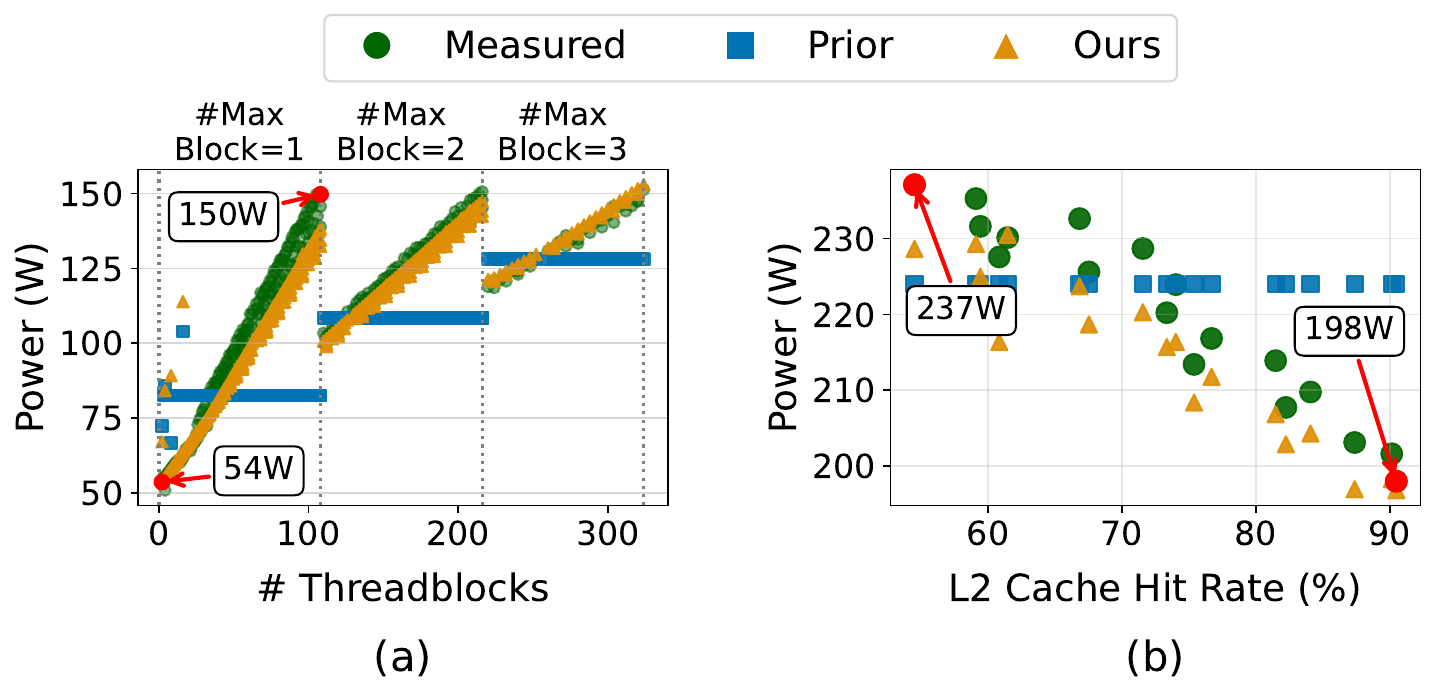}
    \vspace{-6mm}
    \caption{Two cases where kernels with similar latency exhibit different power: (a) Threadblock load imbalance across SMs, and (b) More L2 cache hits reduce power consumption. Adapting prior lightweight models gives inaccurate results, whereas ours produces predictions closely aligned with the measured truth.}
    \vspace{-4mm}
    \label{fig:motivation-power}
\end{figure}
While effective for latency prediction, these models rely on assumptions fundamentally inadequate to provide utilization information. 
Consequently, they fail to capture workloads with similar latency but with vastly different power.

First, consider threadblock load imbalance across SMs. 
When threadblocks distribute unevenly, latency depends on the slowest SM, while power depends on average utilization. 
Microbenchmarks on GEMM kernels (\cref{fig:motivation-power}a) confirm this divergence: power varies by 96 W (54 W vs. 150 W) between workloads with identical maximum threadblocks per SM but different total counts, despite similar latency. 
This distinction is lost if we were to adapt existing lightweight models using the threadblock-level information~\cite{neusight2025} to power estimation, as they do not capture load distribution and would estimate equal power consumption for workloads with the identical maximum threadblocks per SM. 

Second, consider the memory hierarchy.
DRAM accesses consume more energy than L2 cache accesses~\cite{horowitz}.
As a result, workloads with identical kernels and threadblock counts can exhibit different power consumption if their L2 cache hit rates differ. 
Microbenchmarks on GEMM kernels (\cref{fig:motivation-power}b) show a 39 W variation (197 W vs. 236 W) between workloads with identical threadblock counts but different matrix shapes, which affects cache reuse. 
Adapting existing models to power estimation would miss this effect: with memory access modeled as an aggregate footprint without distinguishing the hierarchy, they would predict equal power consumption regardless of the cache hit rates. 
These two cases illustrate that accurate power modeling requires device-wide utilization and memory hierarchy distinction, not just aggregate metrics. 

\begin{table}[t]
    \centering
    \caption{Comparison of Traditional Power Models and Ours}
    \vspace{-1mm}
\begin{tabular}{|c|c|c|c|c|}
\hline
\textbf{\begin{tabular}[c]{@{}c@{}}Obtain\\ Util.\end{tabular}}                & \textbf{\begin{tabular}[c]{@{}c@{}}Walltime\\ for Util.\end{tabular}}                         & \textbf{Power Model}                                                      & \textbf{\begin{tabular}[c]{@{}c@{}}Accuracy\\ (Error \%)\end{tabular}}    & \textbf{\begin{tabular}[c]{@{}c@{}}GPU \\ Arch.\end{tabular}}               \\ \hline \hline
Simulation                                                                     & \begin{tabular}[c]{@{}c@{}}Hours~\cite{pka2021}\end{tabular}                     & \begin{tabular}[c]{@{}c@{}}AccelWattch \\ SASS~\cite{accelwattch2021}\end{tabular}       & \begin{tabular}[c]{@{}c@{}}9.2\% \\ 11\%\\ 13\%\end{tabular}              & \begin{tabular}[c]{@{}c@{}}Volta \\ Pascal\\ Turing\end{tabular}            \\ \hline
                                                                               &                                                                                               & \begin{tabular}[c]{@{}c@{}}AccelWattch\\ HW~\cite{accelwattch2021}\end{tabular}      & 7.5\%                                                                     & Volta                                                                       \\ \cline{3-5} 
                                                                               &                                                                                               & Guerreiro~\cite{guerreiro2018}                                                      & \begin{tabular}[c]{@{}c@{}}7\%\\ 6\%\\ 12\%\end{tabular}                  & \begin{tabular}[c]{@{}c@{}}Pascal\\ Maxwell \\ Kepler\end{tabular}          \\ \cline{3-5} 
\multirow{-3}{*}[0.5cm]{\begin{tabular}[c]{@{}c@{}}Hardware\\ Profiling\end{tabular}} & \multirow{-3}{*}[0.5cm]{\begin{tabular}[c]{@{}c@{}}Minutes~\cite{ncu}\end{tabular}} & Wu~\cite{wu2015}                                                            & 10\%                                                                      & AMD                                                                         \\ \hline
\rowcolor[HTML]{FFFFC7} 
\textbf{\begin{tabular}[c]{@{}c@{}}Lightweight\\ Prediction\end{tabular}}      & \textbf{1.8 sec.}                                                                             & \textbf{\begin{tabular}[c]{@{}c@{}}EnergAIzer \\ (Ours)\end{tabular}} & \textbf{\begin{tabular}[c]{@{}c@{}}8\%\\ 7\%\end{tabular}}        & \textbf{\begin{tabular}[c]{@{}c@{}}Ampere\\ Hopper\end{tabular}}   \\ \hline

\end{tabular}
    
    \vspace{-2mm}
    \label{tab:method-walltime}
\end{table}

\subsection{Addressing the Challenges}

In summary, the core challenge for rapid power estimation is obtaining utilization information.
This work aims to solve this challenge by developing a lightweight performance model that provides utilization information.
Despite their diversity in model architectures (e.g., Transformers~\cite{vaswani2017attention}, convolutions), AI workloads decompose into a common set of kernel types (\cref{fig:background-kernel-breakdown}).
We therefore develop a model targeting these kernels, which employ software optimizations such as tiling and pipelining that create predictable patterns mappable to hardware components. 

Our framework, \name, completes end-to-end power estimation in 1.8 seconds on average, eliminating costly simulations and runtime profiling in traditional power models, while achieving comparably strong accuracy (\cref{tab:method-walltime}). 

\begin{figure}[t]
    \centering
    \includegraphics[width=\linewidth]{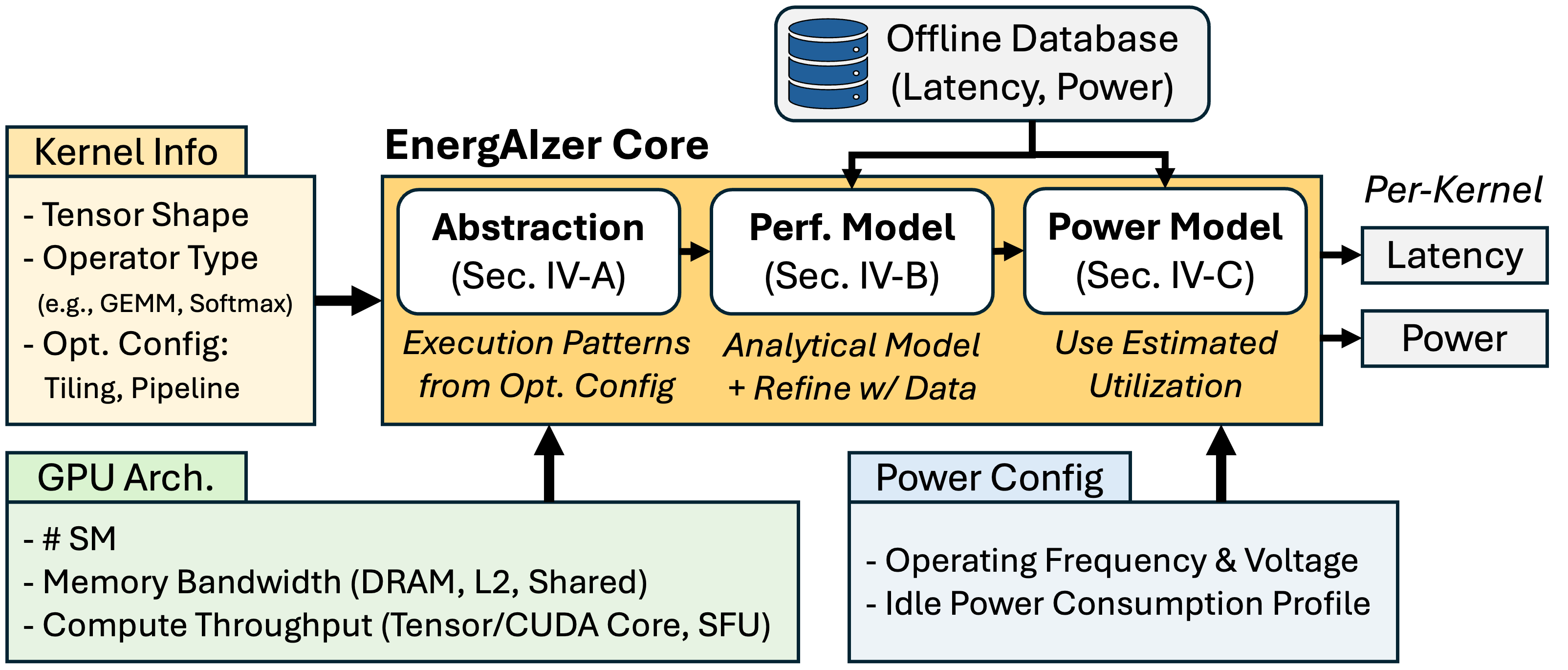}
    \caption{Overview of \name's kernel-level prediction.}
    \vspace{-4mm}
    \label{fig:method-energaizer-core}
\end{figure}

\section{\name Core: Kernel-level Predictions} \label{sec:methods}


\begin{figure*}[t!]
    \centering
    \includegraphics[width=\linewidth]{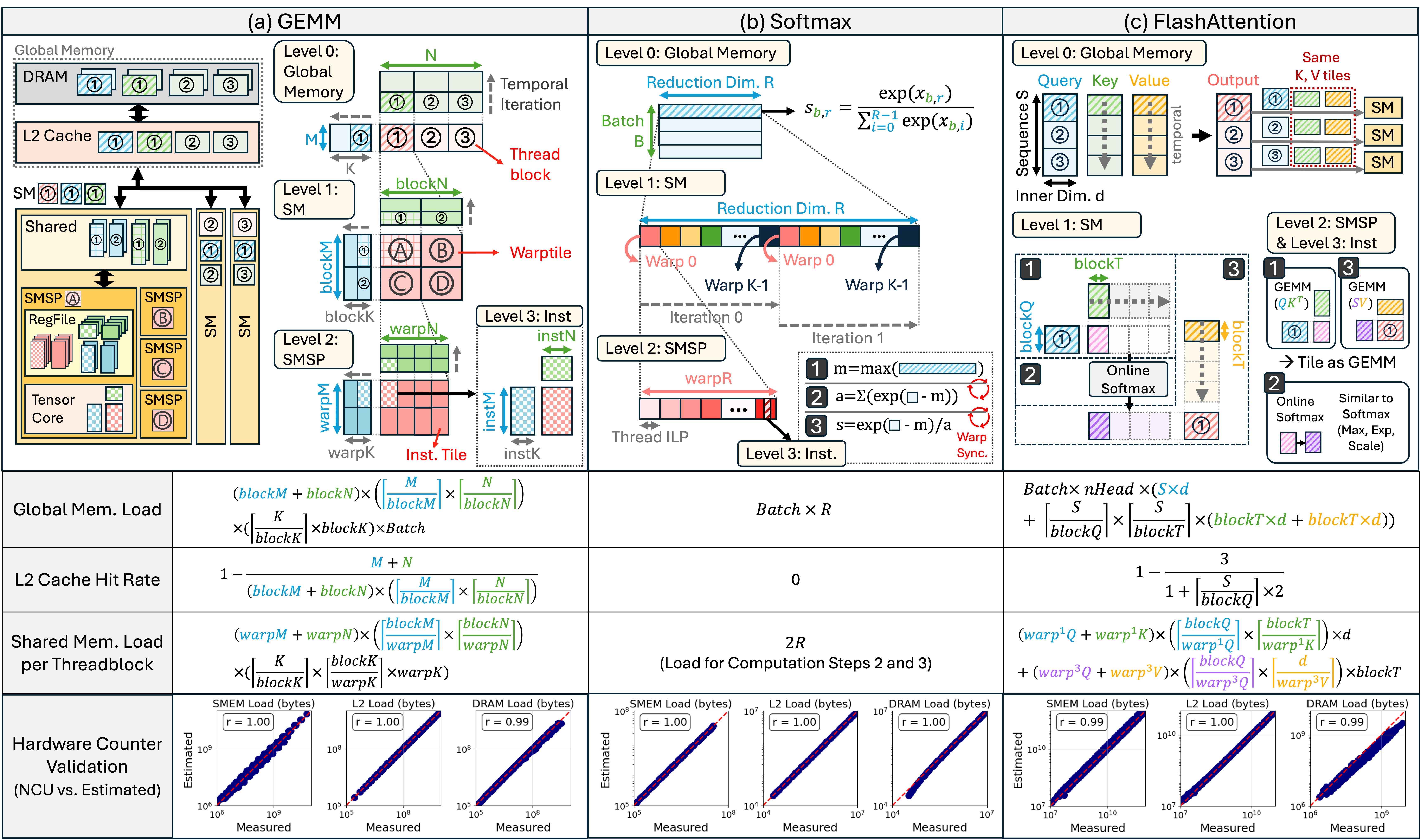}
    \vspace{-2mm}
    \caption{Describing (a) GEMM, (b) Softmax, and (c) FlashAttention kernels' memory traffic and load distributions through their software optimization choices. The analytically derived memory traffic from these choices aligns with the measured traffic.}
    \vspace{-2mm}
    \label{fig:methods-abstraction}
\end{figure*}
In this section, we illustrate \name's kernel-level model predicting latency and power (\cref{fig:method-energaizer-core}). 
As inputs, our kernel-level model takes the tensor shapes and optimization parameters of the kernel, GPU architectural configurations (i.e., the number of SMs, memory bandwidth, and compute throughput), and power configurations like frequency and voltage. 

We develop this model in three steps. 
First, we establish our workload representation (\cref{subsec:method-core-abstraction}).
Software optimization choices, such as tiling, threadblock scheduling, and pipelining, create structured execution patterns that form the basis of our performance model. 
Second, we construct a performance model (\cref{subsec:method-core-perf}) that uses these patterns as a scaffold for empirical data fitting. 
Finally, our power model (\cref{subsec:method-core-power}) uses the predicted utilization to estimate dynamic power. 

\subsection{Workload Representation}\label{subsec:method-core-abstraction}

Dominant kernels in AI workloads, such as GEMM and nonlinear reduction functions, employ a common set of optimizations that analytically determine memory traffic, load distribution across parallel units, and latency hiding behavior. 

\subsubsection{Optimizations}
Tensors are hierarchically partitioned into \emph{tiles} at each level of GPU execution. 
\cref{fig:methods-abstraction}a illustrates tiling for GEMM kernels. 
Threadblock-level tiles determine global memory traffic and work distributions across SMs.
Inside each threadblock, warp-level tiles are distributed across SMSPs and define shared memory traffic.
Finally, instruction-level tiles specify Tensor Core operation granularity. 
Tiles are temporally iterated to accumulate the partial sums (e.g., across the ``K'' dimension in the operand matrices).

\emph{Threadblock swizzling} schedules threadblocks accessing the same input tiles onto nearby SMs, enhancing L2 cache reuse. 
As shown in \cref{fig:methods-abstraction}a, three threadblocks request the same blue input tile.
When threadblocks requesting the same tile execute closely in time, that tile loads once from DRAM while subsequent accesses hit L2 cache.
High-performance GEMM libraries optimize this reuse~\cite{cutlass,cublas}, allowing us to assume ideal behavior: each tile loads once from DRAM, while the L2 cache serves all subsequent accesses. 
This enables the differentiation of DRAM and L2 cache traffic without elaborate cache simulation. 

\emph{Software pipelining} overlaps data movement with computation across temporal iterations (i.e., fetching the next K-dimension tiles in \cref{fig:methods-abstraction}a, for global$\to$shared (G$\to$S) and shared$\to$register (S$\to$R)). 
Pipelining determines exposed latency, essential for performance modeling.

\subsubsection{Beyond GEMM}
These optimizations have been studied in prior analytical models for GEMM kernels~\cite{cutlass,Lym_2019,llmcompass}, with some work also extending to nonlinear reductions~\cite{llmcompass}. 
We systematically extend the analysis to the full set of dominant AI kernel types (i.e., nonlinear, elementwise, and fused kernels), with the specific goal of deriving module-level utilization for power modeling, which is a purpose not addressed by prior analytical models. 

Softmax (\cref{fig:methods-abstraction}b) has a simpler tiling strategy than GEMM. 
Each threadblock handles one row of input, which corresponds to the full reduction dimension. 
Within the threadblock, elements are distributed across warps and processed in three sequential steps: computing the maximum of each row, summing the exponentials, and producing the final output. 
The first step loads data into shared memory, which is reused by the two subsequent steps. 


FlashAttention~\cite{dao2022flashattention} is a widely adopted fused kernel for self-attention, combining two GEMM operations and an online softmax into a single pass to reduce memory traffic.
Despite this complexity, it preserves the same optimization structure (\cref{fig:methods-abstraction}c). 
At the threadblock level, the output and Query tensors are tiled along the sequence dimension and distributed across SMs, while the Key and Value tensors form the inner temporal iteration dimension, analogous to the K-dimension in GEMM. 
Multiple threadblocks access the same Key and Value tiles, creating threadblock swizzling opportunities as in GEMM. 
Within each threadblock, the three computation steps (two matrix multiplications and a softmax) mirror the warp and instruction-level execution patterns of GEMM and softmax. 


\subsubsection{Validation}
We validate that tiling parameters and the ideal threadblock swizzling determine memory traffic. 
We analytically derive the total load traffic for shared memory, L2 cache, and DRAM (detailed equations shown in \cref{fig:methods-abstraction}) and compare against the NCU-profiled hardware counter values from NVIDIA A100-40GB-PCIE GPU. 
Across 790+ GEMM\footnote{cuBLAS \texttt{float16} GEMM, Batch$\in[2,128]$, M,N,K$\in[128,4096]$}, 70 Softmax\footnote{PyTorch \texttt{cunn\_SoftMaxForwardSmem}}, and 380+ FlashAttention\footnote{PyTorch \texttt{scaleddotproductattention} using FlashAttention} kernels, we observe near-perfect correlations ($r\geq0.99$). 
For FlashAttention, the shared memory traffic is derived from the warp-level tiling of its two constituent GEMM steps.
\subsection{Performance Modeling}\label{subsec:method-core-perf}

\begin{figure}[t]
    \centering
    \includegraphics[width=\linewidth]{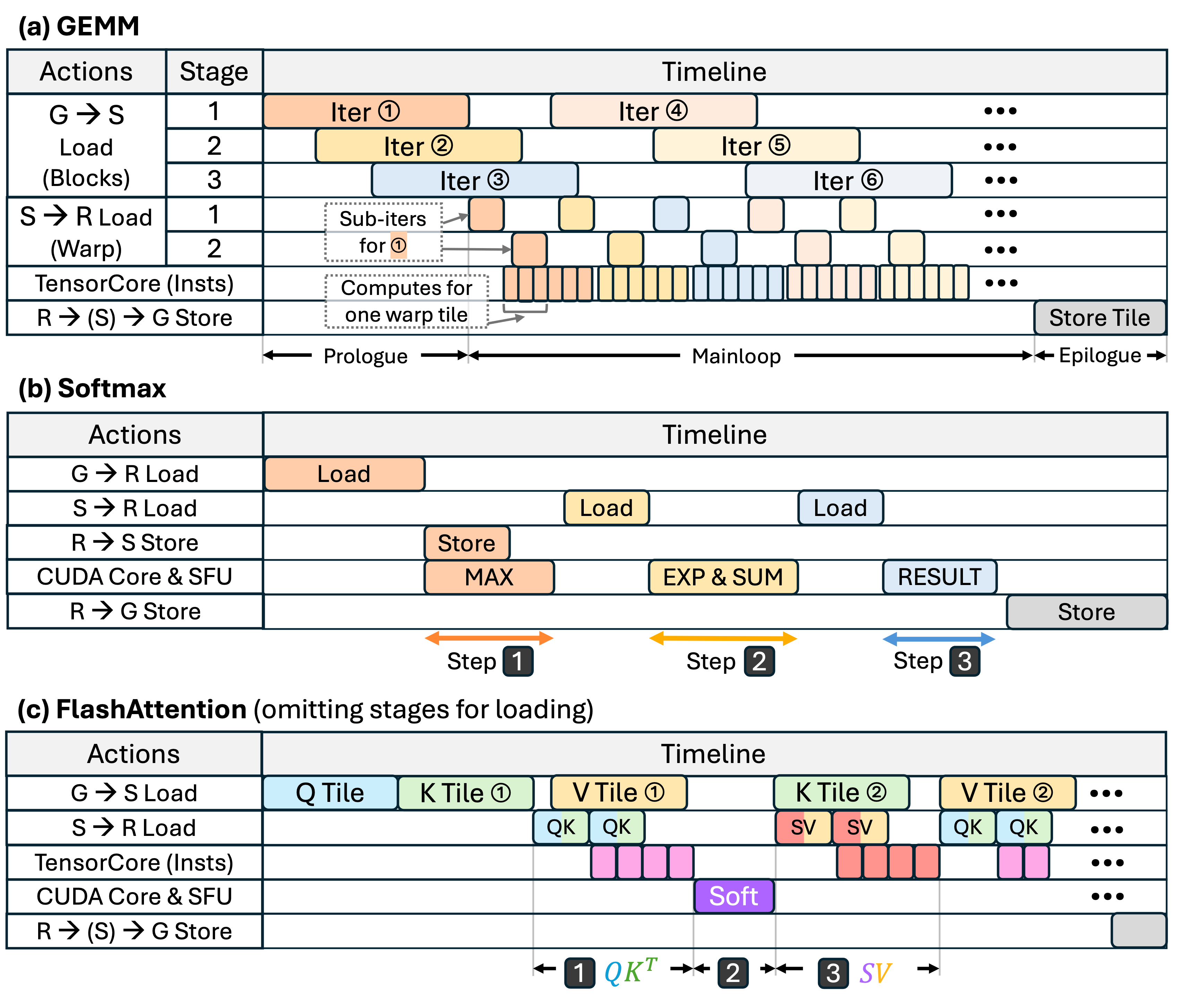}
    \vspace{-2mm}
    \caption{Illustration of timelines for (a) GEMM, (b) Softmax, and (c) FlashAttention kernels. For warp-level and instruction-level actions, only the timeline of a single warp is shown for simplicity.}
    \vspace{-4mm}
    \label{fig:method-perf-timeline}
\end{figure}
\subsubsection{Timeline Construction}

Our performance model constructs an execution timeline of coarse-grained actions, where tiling defines the granularity of actions (e.g., data load/store amount, compute instruction counts) and pipelining determines their overlap reflecting dependencies. 
This timeline forms the analytical scaffold that exposes module-level utilization. 

We illustrate this construction using a GEMM kernel (\cref{fig:method-perf-timeline}a), focusing on a single threadblock. 
In this example, three pipeline stages are used for G$\to$S loads (note that the number of stages can vary across kernels). 
This multi-stage pipeline allows one stage's G$\to$S load to overlap with two other stages' lower-level operations (S$\to$R load and Tensor Core compute), effectively hiding the latency of global memory access. 
Similarly, S$\to$R loads use double-buffering to hide the latency of shared memory access.
After iterating across all K-dimension tiles, the kernel writes the result back to global memory, optionally staging through shared memory for better coalesced stores (i.e., R$\to$S then S$\to$G). 

This timeline construction generalizes to other kernel types.
For Softmax(\cref{fig:method-perf-timeline}b), each threadblock loads its tile from global memory to registers (G$\to$R).
While computing the maximum values, this tile is stored to shared memory (R$\to$S) and reused in the two subsequent computation steps (S$\to$R), avoiding redundant global memory accesses. 
FlashAttention (\cref{fig:method-perf-timeline}c) closely mirrors GEMM: Key and Value tile loading forms a natural double-buffering pattern analogous to GEMM's G$\to$S pipelining, and three computation steps follow the warp and instruction-level patterns of GEMM and softmax. 
Other kernel types, such as elementwise operations, are modeled similarly from their tiling and pipelining parameters, though we omit their detailed description here.

Finally, this single-threadblock timeline extends to device-wide execution.
Each SM processes multiple concurrent threadblocks, and we assume SM resources, such as shared memory bandwidth and Tensor Core throughput, are divided equally among them. 
A kernel then executes in waves, where each wave corresponds to one top-level iteration of the timeline across all concurrent threadblocks. 
When threadblocks are distributed unevenly across SMs, some SMs process an additional threadblock, which we call ``busy'' SMs, while the rest are ``lazy''. 
We use the busy SM's timeline for latency prediction and a weighted average across busy and lazy SMs for utilization estimation.

\subsubsection{Latency Predictions}
Having established the timeline structure, we now describe how action latencies are calculated.
As the first-order approach, an action's ideal latency for a given module follows:
\begin{equation}
    t_\text{action}^\text{module} = \frac{\text{amount of work}}{{\text{bandwidth}}/{\text{concurrency}}}
\end{equation}
where the amount of work represents data transfer size or compute FLOPs. 
Since each action engages multiple modules simultaneously, the action's latency is determined by the slowest participating module. 
For example, a G$\to$S load involves DRAM, L2 cache, and shared memory, giving $t_{G\to S}=\max(t^{DRAM}_{G\to S}, t^{L2}_{G\to S}, t^{shared}_{G\to S})$.

These individual action latencies are then composed into the total execution time, reflecting pipelining. 
When multiple actions are overlapped due to pipelining, the exposed latency is determined by the slowest one. 
In the three-stage pipelining example in \cref{fig:method-perf-timeline}a, one G$\to$S load is overlapped with two stages of lower-level operations, and the exposed latency is the maximum of these two terms. 

\begin{figure}[t]
    \centering
    \includegraphics[width=\linewidth]{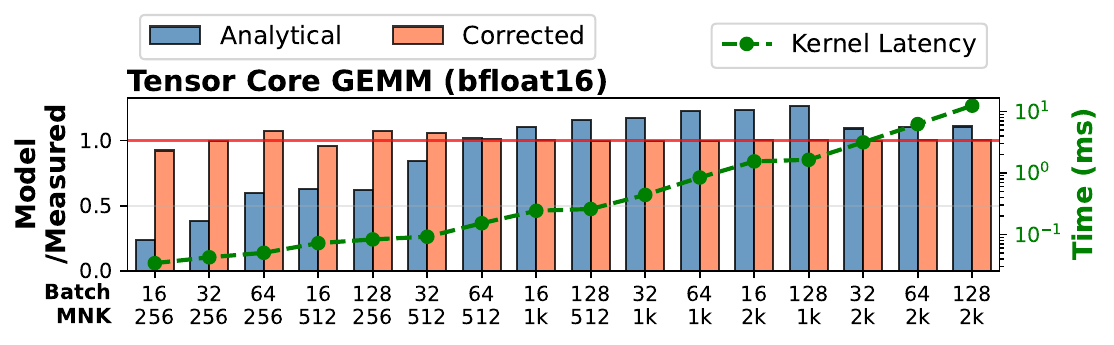}
    \includegraphics[width=\linewidth]{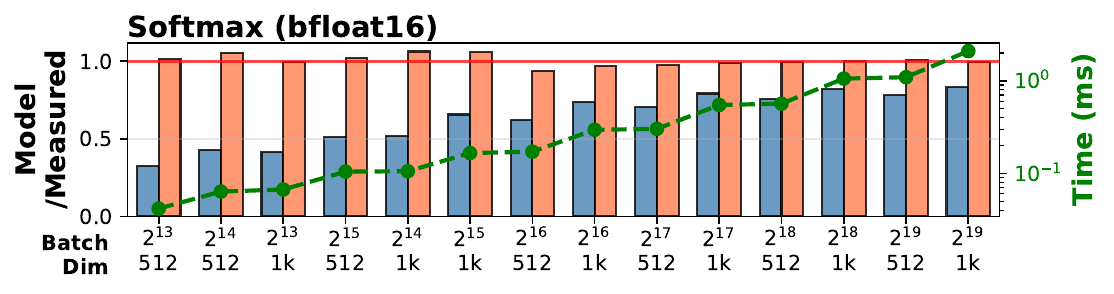}
    \includegraphics[width=\linewidth]{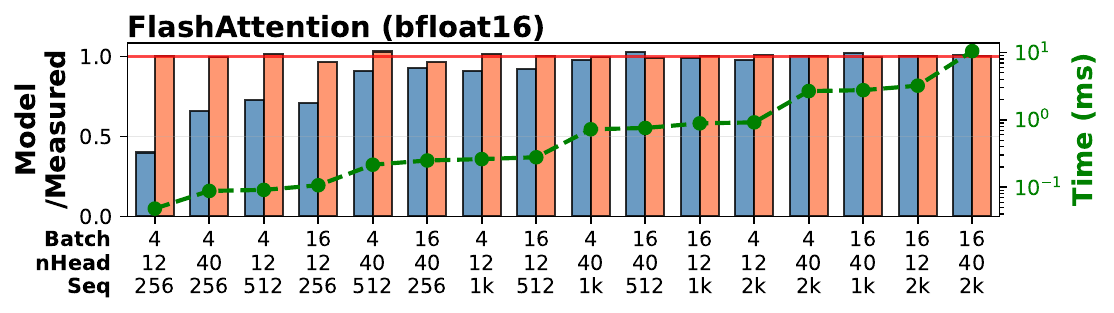}
    \vspace{-2mm}
    \caption{Estimation accuracy of the analytical model and the corrected prediction across diverse workload shapes (x-ticks indicate workload tensor sizes).}
    \vspace{-2mm}
    \label{fig:methods-perf-analytical-corrected}
\end{figure}
\noindent\textbf{Analytical Model Validation and Limitations~}
The timeline construction with ideal bandwidth-based latency modeling captures the dominant execution patterns from tiling and pipelining.
We evaluate this analytical model's latency predictions on GEMM, Softmax, and FlashAttention across diverse tensor shapes (\cref{fig:methods-perf-analytical-corrected}).
For large tensors with long latency, it alone achieves $\leq20\%$ error.
However, for small tensors, the latency is severely underestimated with error reaching up to $80\%$.
This gap reveals that the analytical model misses secondary effects, such as kernel initialization, control-flow overheads (e.g., calculating loop indices, warp scheduling), and potential memory bank conflicts, that become proportionally significant for short kernels.


\noindent\textbf{Empirical Refinement~}
Modeling these secondary effects based on instruction-level analysis would sacrifice our scalability objective.
Instead, following recent lightweight performance models~\cite{li2023micro,neusight2025,habitat2021} that combine coarse-grained abstraction with data fitting, we apply empirical corrections that refine the latency prediction:
\begin{equation}
    \hat{t}_\text{corrected} = \lambda \times t_\text{ideal} + \varepsilon
\end{equation}
where $\lambda$ captures effective bandwidth degradation potentially from scheduling overheads and memory access inefficiencies, while $\varepsilon$ represents fixed costs independent of the workload size, such as kernel launch overhead. 

Since pipelining behavior differs across execution phases, we apply separate empirical corrections to each phase rather than a single global term. 
GEMM kernels naturally divide into three phases: (1)~the prologue fills the pipeline before starting any computation, (2)~the mainloop iterates over K-dimension tiles with pipelining partially or fully hiding data transfer latency, and (3)~the epilogue stores results without overlap.
As each phase exhibits distinct latency-hiding characteristics, we fit a separate $\lambda$ per phase (e.g., $\lambda_p$ for prologue, $\lambda_m$ for mainloop, and $\lambda_e$ for epilogue) and one global $\varepsilon$ term:
\begin{equation}
    \hat{t}_\text{corrected}^\text{GEMM} = \lambda_p\cdot t_\text{p, ideal} + \lambda_m\cdot t_\text{m, ideal} + \lambda_e\cdot t_\text{e, ideal} + \varepsilon
\end{equation}
FlashAttention follows a similar three-phase structure, with the loading of the Query and the first Key tile forming the prologue. 
Nonlinear and elementwise kernels lack explicit pipelining and are therefore corrected with a single $\lambda$ term applied to the total latency. 

\noindent\textbf{Coefficient Fitting~}
We determine these coefficients by fitting them to an offline database of kernels with measured latencies. 
For each kernel group sharing the same tiling and pipelining strategy, we solve a separate quadratic optimization that minimizes the prediction error for the samples in that group:
\begin{equation}
    \min_{\lambda, \varepsilon} \sum_{i\in\text{group}} ||\hat{t}_{i,\text{corrected}} - t^*_{i,\text{measured}}||^2
\end{equation}
where $t^*_{i,\text{measured}}$ is the measured latency of the sample $i$. 

Despite its simplicity, this linear correction is effective.
Errors reduce to $<$5\% across both small and large kernels (\cref{fig:methods-perf-analytical-corrected}), demonstrating that lightweight empirical refinement successfully addresses the corner cases while maintaining scalability. 



\subsubsection{Deriving Utilization}
From the constructed timeline, we extract module-level utilization for the six key modules explicitly modeled: DRAM, L2 cache, shared memory, Tensor Cores, CUDA Cores (for regular floating-point arithmetic), and Special Function Units (for exponentials and other nonlinear functions). 
For each module, utilization is computed as the ratio of its active time to the total kernel latency. 
A module's active time is obtained by summing the latencies of all actions that engage it, applying the phase-specific $\lambda$ correction to each action according to its execution phase. 
For instance, G$\to$S loads contribute to the active time of DRAM, L2 cache, and shared memory, while the matrix multiplication instructions contribute to Tensor/CUDA Cores.
For device-wide estimation, we compute per-SM utilization for each module and take a weighted average across busy and lazy SMs. 


\subsection{Power Modeling}\label{subsec:method-core-power}
With module-level utilization derived from the performance model, we now estimate dynamic power consumption using the standard equation: 
\begin{equation}
    P^\text{dyn} = \alpha_\text{DRAM} \cdot C_\text{D}V_\text{D}^2f_\text{D} + \sum_{\text{modules}} \alpha_\text{module}\cdot C_\text{m} V^2f
    \label{eq:method-dynamic-power}
\end{equation}
where $\alpha$ represents utilization, $V$ and $f$ are the supply voltage and frequency (provided as DVFS configuration inputs), and $C$ is the hardware-specific parameter that we determine from empirical data fitting. 
Note that DRAM is an off-chip component with its own frequency ($f_\text{D}$) and voltage ($V_\text{D}$) domain.
This approach aligns with traditional power modeling~\cite{gpuwattch2013,accelwattch2021,ipp2010,guerreiro2018} in using empirical fitting for finding $C$, but differs in how utilization $\alpha$ is obtained. 
Similar to the performance model, $C$ coefficients are fitted using quadratic programming ($\min_{C}\sum_{i\in\text{group}} ||P_{i}^{\text{dyn}} - P^{*{\text{dyn}}}_{i,\text{measured}}||^2$), grouping the kernels by their tiling and pipelining configurations and fitting within each group. 
Total power consumption combines the estimated dynamic power with the idle power measured at the frequency $f$. 

Our offline database includes power measurements across multiple operating frequencies.
We fit $C$ coefficients to minimize the error across the frequency range, enabling power estimation at any frequency within the range without additional measurements during the inference.

\begin{figure}[t]
    \centering
    \includegraphics[width=0.9\linewidth]{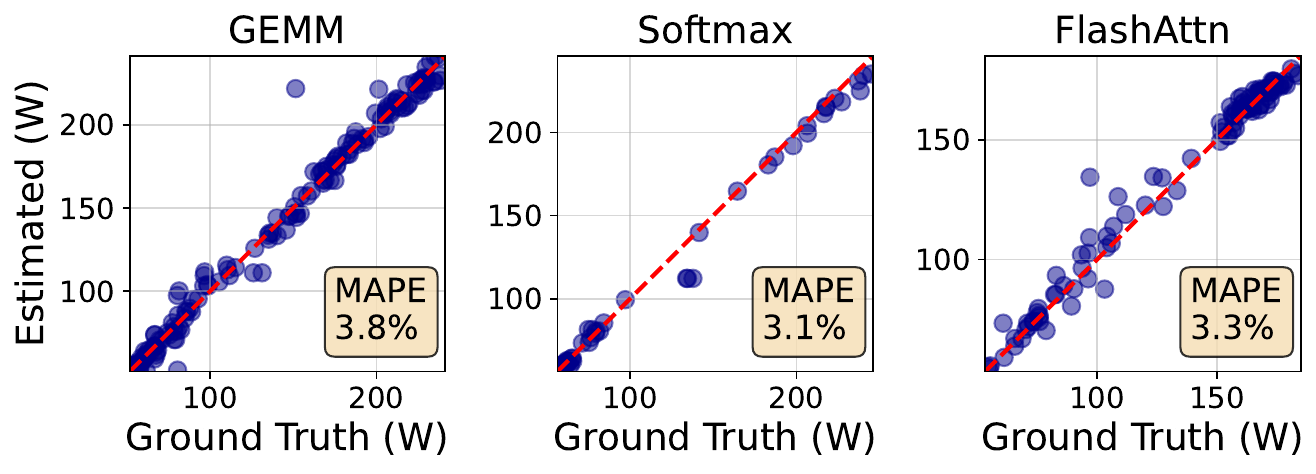}
    \caption{Validating power estimation for GEMM, Softmax, and FlashAttention kernels, along with the mean average percentage error (MAPE) annotated.}
    \vspace{-2mm}
    \label{fig:methods-power-experiments}
\end{figure}

We validate the kernel-level power predictions for diverse kernel types benchmarked on A100-40GB-PCIE GPU at 900 MHz as a representative frequency  (\cref{fig:methods-power-experiments}).
Our model achieves an average error of 3.1-3.8\% across these kernel types. 

\section{\name Framework: End-to-End Predictions} \label{sec:framework}
\begin{figure}[t]
    \centering
    \includegraphics[width=0.9\linewidth]{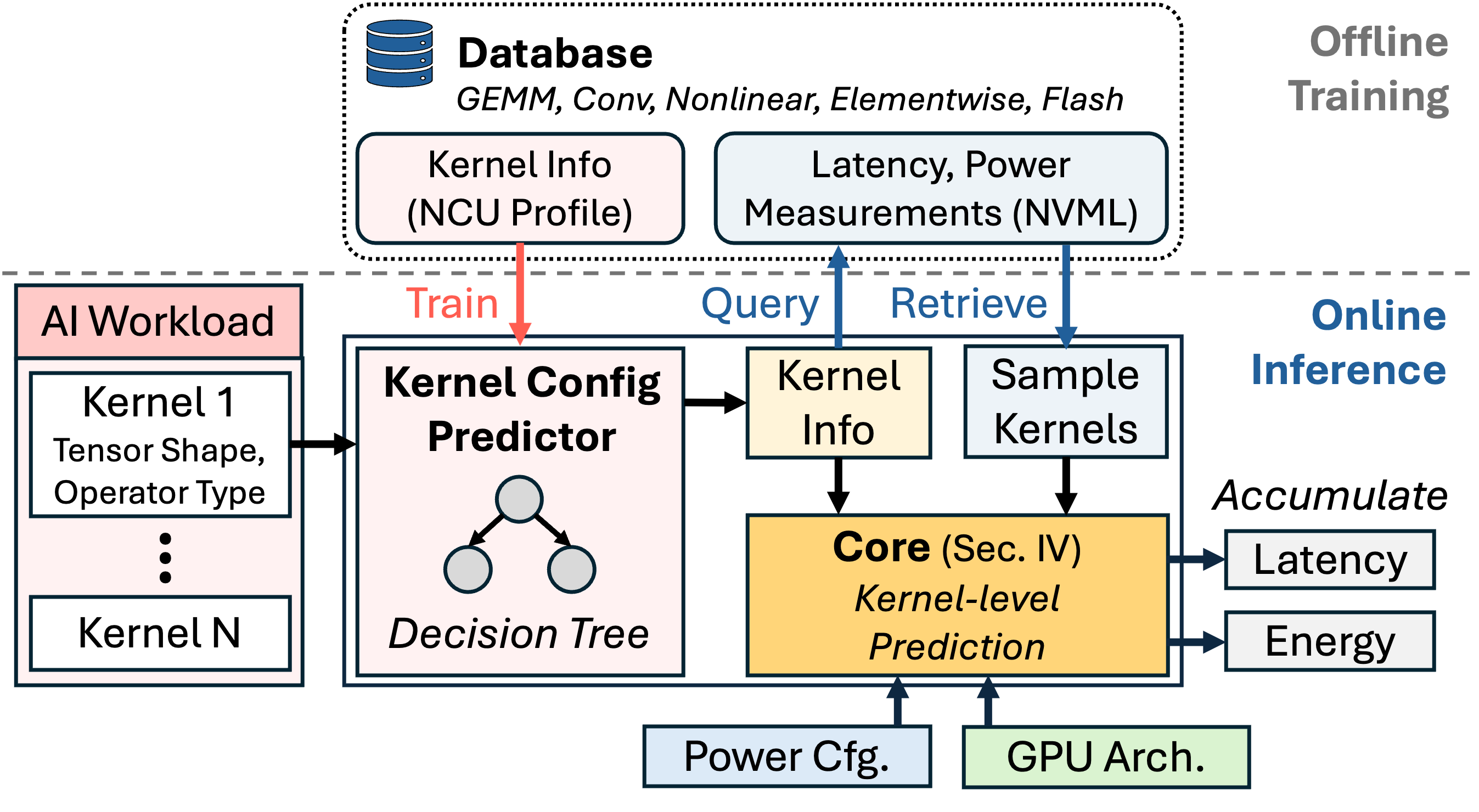}
    \caption{\name system overview, including the frontend kernel configuration predictor and offline-collected database.}
    \label{fig:methods-framework}
\end{figure}

\cref{fig:methods-framework} illustrates the end-to-end prediction workflow of \name. 
The framework takes three inputs: a sequence of operators (e.g., GEMM, nonlinear, elementwise) comprising the AI workload, GPU architecture configurations (e.g., number of SMs, memory bandwidth, and compute throughput), and power configurations (operating frequency and voltage). 
For each operator, \name predicts its kernel-level latency and power using the models described in \cref{sec:methods}, and accumulates these predictions to produce the end-to-end latency and average power consumption for the queried AI workload. 
Additionally, users can configure the GPU architecture and power parameters to explore the design spaces and DVFS schemes without requiring physical hardware. 


A key requirement for kernel-level prediction is knowing each operator's tiling and pipelining parameters. 
Requiring users to provide these directly would necessitate offline profiling, undermining the framework's scalability goal. 
Instead, \name uses a decision tree trained offline to predict these parameters from the operator's input and output tensor shapes. 
To train this decision tree, we extract the tiling and pipelining parameters from kernels in our offline database. 
For high-performance GEMM libraries (cuBLAS~\cite{cublas}, CUTLASS~\cite{cutlass}), these parameters are encoded in the kernel name, from which we parse threadblock tiles, pipeline stages, and instruction shapes directly. 
For nonlinear and elementwise operators, which use open-source PyTorch kernels, the parameters are extracted from the source code. 
The decision tree then learns to map tensor shapes to tiling and pipelining strategies for each operator type, achieving 93\% prediction accuracy for Tensor Core GEMM kernels.


\name separates one-time offline training from online inference. 
During offline training, latency and power measurements are collected through the NVML~\cite{nvml} interface on the training GPU, along with kernel information (i.e., kernel name, grid/block sizes, and concurrency) obtained via NCU~\cite{ncu} for decision tree training. 
This offline collection is a one-time upfront cost. 
During inference, \name requires no additional data collection or profiling, providing rapid predictions.

\section{Results} \label{sec:results}

\begin{table}[t!]
\centering
\caption{Offline Database Used for Experiments}
\vspace{-1mm}
\footnotesize
\begin{tabular}{|lll|}
\hline
\multicolumn{2}{|l|}{\textbf{Training GPUs}}                                      & NVIDIA A100-40GB-PCIE, A10                                                                                        \\ \hline
\multicolumn{2}{|l|}{\textbf{Collected Data}}                                     & \begin{tabular}[c]{@{}l@{}}Latency, power,\\ kernel optimization configurations \end{tabular} \\ \hline
\multicolumn{2}{|l|}{\textbf{Frequency Range}}                                    & 4 frequencies between 510-1410 MHz                                                                                \\ \hline \hline
\multicolumn{3}{|c|}{\textbf{Collected Kernel Information}}                                                                                                                                           \\ \hline
\multicolumn{1}{|l|}{\textbf{Type}}           & \multicolumn{1}{l|}{\textbf{Library}}      & \textbf{Sample Count \& Precision}                                                                                         \\ \hline
\multicolumn{1}{|l|}{GEMM}           & \multicolumn{1}{l|}{{cuBLAS}}       & 3300+ shapes; \texttt{bf16}, \texttt{fp32}                                                                                   \\ \hline
\multicolumn{1}{|l|}{Convolution}    & \multicolumn{1}{l|}{PyTorch} & 3000+ shapes; \texttt{bf16}                                                                                            \\ \hline
\multicolumn{1}{|l|}{Nonlinear}      & \multicolumn{1}{l|}{PyTorch}      & 450+ shapes; \texttt{bf16}, \texttt{fp32}                                                                                    \\ \hline
\multicolumn{1}{|l|}{Elementwise}    & \multicolumn{1}{l|}{PyTorch}      & 2200+ shapes; \texttt{bf16}, \texttt{fp32}                                                                                   \\ \hline
\multicolumn{1}{|l|}{FlashAttention} & \multicolumn{1}{l|}{PyTorch}      & 500+ shapes; \texttt{bf16}                                                                                             \\ \hline
\end{tabular}
\vspace{-2mm}
\label{tab:result-database}
\end{table}

\begin{figure*}[t]
    \centering
    \includegraphics[width=\linewidth]{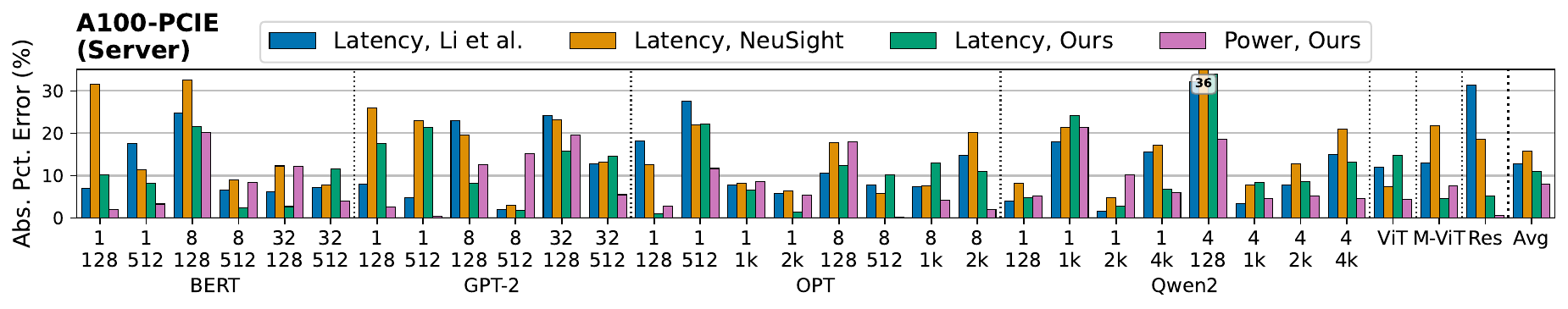} 
    \includegraphics[width=\linewidth]{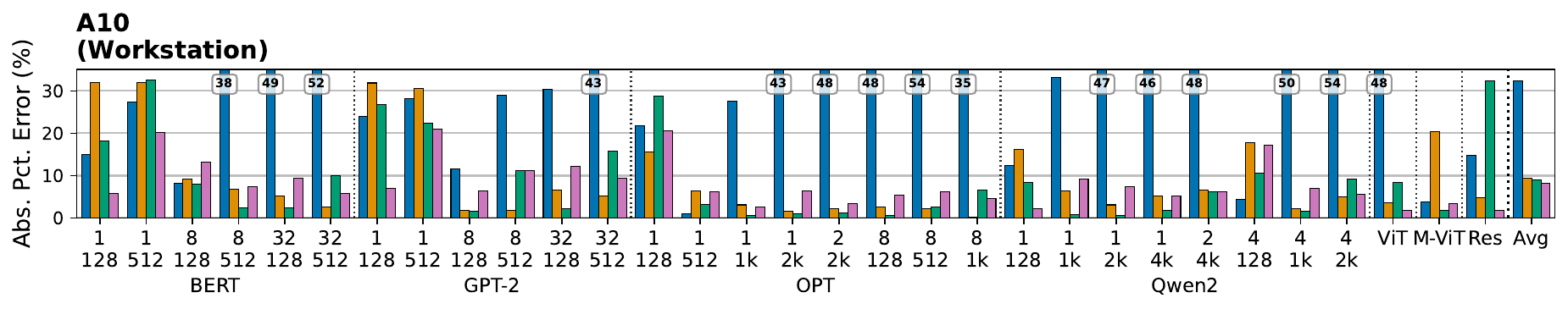}
    \caption{End-to-end latency and power estimation errors for NVIDIA A100-40GB-PCIE and A10 GPUs, with the operating frequency of 900 MHz. For language workloads, the batch and sequence length are shown in the x-tick. Errors higher than 40\% are clipped.}
    \vspace{-2mm}
    \label{fig:results-end-to-end}
\end{figure*}

In this section, we demonstrate \name's prediction capabilities and its applications for exploring diverse design choices. 
First, we show both the latency and power estimation accuracy for AI workloads (\cref{sec:results-same-gpu}). 
Then, we demonstrate \name's utility for exploring voltage-frequency scaling (\cref{sec:results-dvfs}), GPU architecture configurations (\cref{sec:results-extrapolation}), and algorithm-level choices (\cref{sec:results-fusion}).

Common to all experiments, we collect an offline kernel database on NVIDIA A100-40GB-PCIE and A10 GPUs, covering several tensor shapes for major operator types, to train \name (\cref{tab:result-database}). 

\subsection{Latency \& Power Estimation Accuracy} \label{sec:results-same-gpu}

\noindent\textbf{End-to-end~}
\cref{fig:results-end-to-end} shows end-to-end latency and power estimation accuracy across diverse language (BERT-Large~\cite{bert}, GPT-2~\cite{gpt2}, OPT-1.3B~\cite{zhang2022opt}, Qwen2-1.5B~\cite{yang2024qwen2}) and vision (ResNet101~\cite{resnet}, ViT~\cite{vit}, MobileViT~\cite{mehta2021mobilevit}) workloads with varying batch sizes and sequence lengths.
Note that language workloads are executed in eager mode (i.e., FlashAttention is disabled) for this experiment. 
\name achieves 11.0\% latency and 8.0\% power error on the server-grade A100-40GB-PCIE, and 8.8\% latency and 8.2\% power error on the workstation-grade A10, averaged across all workloads. 
Latency estimation is competitive with state-of-the-art lightweight performance models (Li et al.~\cite{li2023micro}, NeuSight~\cite{neusight2025}), while additionally providing power estimates that these prior models do not offer.

\name completes both latency and power estimation in 1.8 seconds per workload on average\footnote{At Intel i5-1345U CPU. No GPU is required for inference.}.
For language workloads, the per-workload walltime ranges from 1.1 to 2.8 seconds. 
Compared to 452 to 8192 seconds required for NCU profiling to obtain hardware counters, we achieve a speedup of 317 to 3856$\times$. 
Note that prior GPU power models are limited to older architectures (e.g., Volta~\cite{accelwattch2021}) and rely on simulation or profiling with impractical walltimes. 
\name offers power modeling for recent GPU generations with orders of magnitude faster prediction times. 

\begin{figure}[t!]
    \centering
    
    \subfloat[Kernel-level Latency Estimation Accuracy]{
    \includegraphics[width=\linewidth]{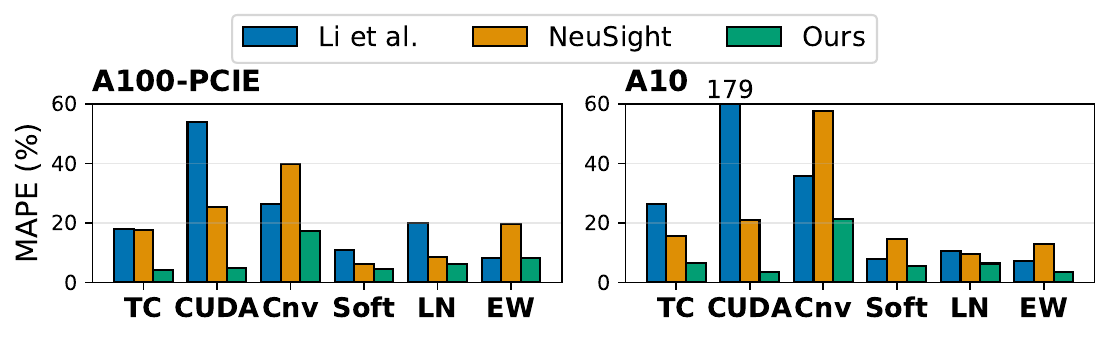}
    }

    \subfloat[Latency Estimation Error Distribution for \texttt{bf16} GEMM]{
    \includegraphics[height=2.7cm, valign=t]{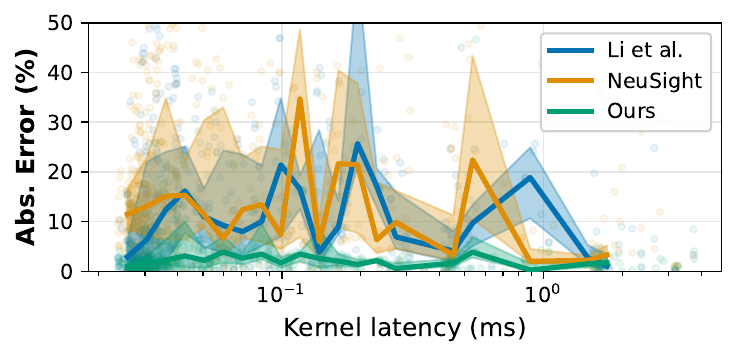} \quad
    \includegraphics[height=2.4cm, valign=t]{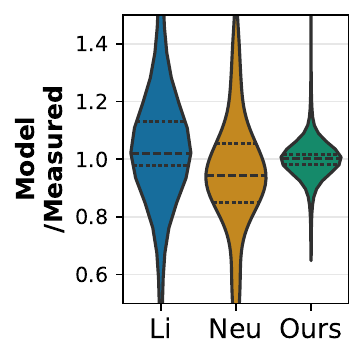}
    }
    \vspace{-1mm}
    \caption{Kernel-level latency estimation with detailed comparisons against the state-of-the-art performance models. The abbreviations stand for: TC - Tensor Core GEMM (\texttt{bf16}), CUDA - CUDA Core GEMM (\texttt{fp32}), Cnv - Convolution, Soft - Softmax, LN - Layer Normalization, EW - Elementwise.}
    \vspace{-2mm}
    \label{fig:results-single-op}
\end{figure}

\noindent\textbf{Kernel-level~}
We take a deeper look into the error distribution at the kernel level (\cref{fig:results-single-op}a), evaluating on a held-out test set randomly split from our offline database. 
\name achieves 4-20\% mean error across kernel types.
A notable improvement is for Tensor Core GEMM kernels, where our framework achieves 4\% mean error compared to 18\% for prior lightweight models.
As shown in the error distribution across kernel sizes (\cref{fig:results-single-op}b), prior models suffer on small kernels with short execution times, where secondary effects like fixed costs from kernel initialization become proportionally significant. 
Our correction terms account for these effects, maintaining consistently low error across the kernel sizes. 

\noindent\textbf{Sources of Error~}
We analyze the sources of estimation error at two levels.
At the single kernel level, the first source of error is the misprediction of tiling configurations by the decision tree.
For GEMM kernels, this accounts for the worst-case latency errors: assuming a tile oracle with perfect configuration predictions, the worst-case error reduces from 23\% to just 1.4\%. 
This misprediction also propagates to end-to-end errors for AI workloads.
For the worst-case outliers in \cref{fig:results-end-to-end}, a tile oracle can reduce latency error (e.g., 24\%$\to$11\% for Qwen2, Batch=1, Seq=1k; 34\%$\to$28\% for Qwen2, Batch=4, Seq=128 for A100). 
The second source is the objective function for the coefficient fitting, which minimizes absolute errors uniformly across the sample kernels.
This results in larger relative errors for smaller workloads, where the same absolute deviation represents a higher percentage of the total. 

At the end-to-end workload level, an additional source of error is CPU kernel launch overhead. 
For small nonlinear and elementwise kernels, this overhead becomes comparable to the actual GPU execution time. 
While negligible for large workloads, its aggregate contribution across many small kernels can increase end-to-end errors. 

\noindent\textbf{Sensitivity to Database~}
To verify that the empirically fitted coefficients do not overfit to specific kernels in the database, we conduct a sensitivity analysis by randomly subsampling 50\% of the database kernels and re-evaluating end-to-end predictions across all workloads in \cref{fig:results-end-to-end}. 
Across three independent trials, we observe minimal degradation and low variance in errors (latency: 10.9-12.3\%, power: 8.1-9.3\%), showing that the fitted coefficients generalize well beyond the specific kernels used for calibration. 

\begin{figure}[t]
    \centering
    \includegraphics[width=0.9\linewidth]{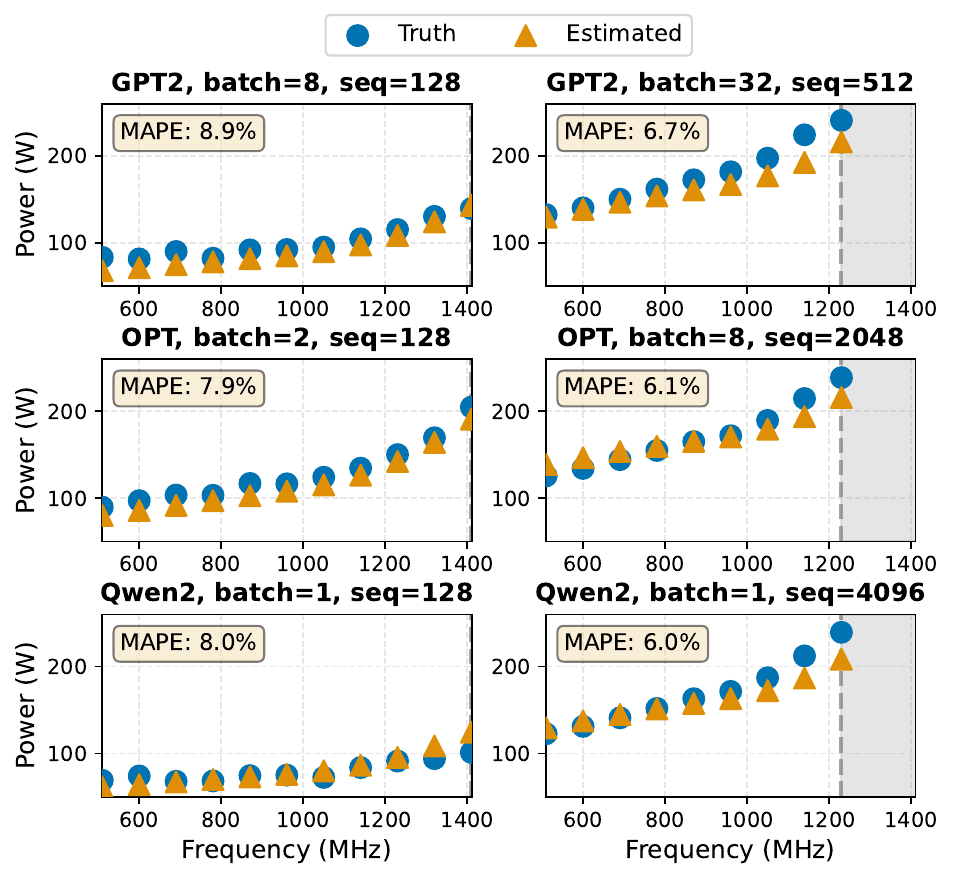}
    \caption{Power estimations across the frequency range of 510-1410 MHz at A100-40GB-PCIE.}
    \vspace{-2mm}
    \label{fig:results-dvfs}
\end{figure}
\subsection{Exploration: Voltage-Frequency Scaling}\label{sec:results-dvfs}
Voltage-frequency scaling is a commonly used technique for energy management, which can benefit from accurate power prediction across operating points. 
We evaluate \name's ability to estimate power consumption across frequencies (510-1410 MHz) on A100-40GB-PCIE. 
We only modify the power configuration parameters provided as an input to \name, including the target frequency and voltage, and idle power at that frequency. 
\cref{fig:results-dvfs} shows the measured ground truth and estimated power consumption.
Our framework captures distinct scaling behaviors of low-utilization (small batch/sequence, left panel) and power-capped (large batch/sequence, right panel) workloads, achieving 6-9\% MAPE across frequencies. 


\begin{table}[t]
    \footnotesize
    \centering
    \caption{Specifications of GPU Architecture Config.}
    \begin{tabular}{|l|c|c|c|c|}
\hline
                   & \textbf{\begin{tabular}[c]{@{}c@{}}A100\\ PCIE\end{tabular}} & \textbf{\begin{tabular}[c]{@{}c@{}}A100\\ SXM\end{tabular}} & \textbf{H100}             & \textbf{L40S} \\ \hline \hline
\#SM               & 108                                                          & 108                                                         & 132                       & 142           \\ \hline
BF16 TFLOPS        & 312                                                          & 312                                                         & 989.4                     & 362.05        \\ \hline
DRAM Size (GB)     & 40                                                           & 80                                                          & 80                        & 48            \\ \hline
DRAM BW (GB/s)     & 1555                                                         & 2038                                                        & 3352                      & 846           \\ \hline
DRAM Freq. (MHz)   & 1215                                                         & 1539                                                        & 2619                      & 9000          \\ \hline
DRAM Spec.         & \multicolumn{1}{l|}{HBM2}                                    & \multicolumn{1}{l|}{HBM2e}                                  & \multicolumn{1}{l|}{HBM3} & GDDR6         \\ \hline
TDP (W)            & 250                                                          & 400                                                         & 700                       & 350           \\ \hline
Target Freq. (MHz) & -                                                            & 1410                                                        & 1830                      & 1920          \\ \hline
\end{tabular}
\vspace{-2mm}
    \label{table:results-gpu-arch}
\end{table}
\begin{figure}[t]
    
    \centering
    \includegraphics[width=\linewidth]{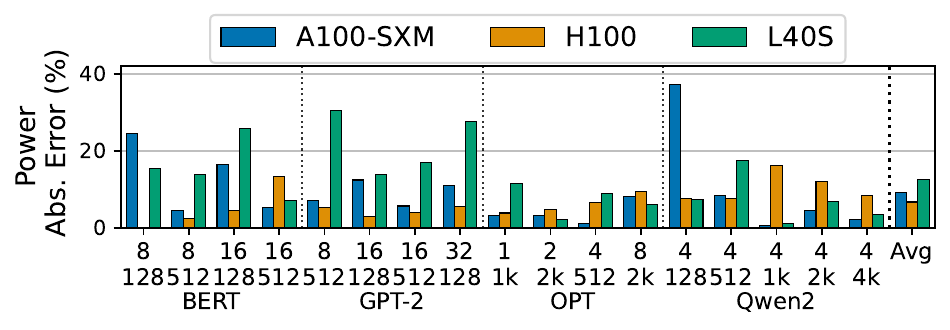}
    \caption{Power estimation errors of forecasting for new GPU configurations.}
    \vspace{-2mm}
    \label{fig:results-extrapolation}
\end{figure}
\subsection{Exploration: GPU Architecture Configuration} \label{sec:results-extrapolation}

Our framework enables exploring GPU architectural configurations by adapting parameters, such as SM count, memory bandwidth, and compute throughput, as inputs.
This allows forecasting power for new GPU architectures without requiring data collection from those target designs. 
We evaluate two scenarios, exploring within the same GPU generations and across the generations, with target GPU configurations summarized in \cref{table:results-gpu-arch}. 
First, within the Ampere generation, we predict power for A100-80GB-SXM using only the database collected from A100-40GB-PCIE, achieving 9.1\% average power error. 

Second, for cross-generation exploration, we evaluate H100 (Hopper) and L40S (Lovelace), using the database collected from Ampere-generation GPUs, achieving 6.7\% and 12.7\% power error, respectively. 
The higher error for L40S likely stems from a difference in memory technology.
\name assumes consistent energy efficiency (pJ/bit and pJ/MAC) across GPUs.
However, unlike the HBM-based A100 and H100, L40S uses GDDR6, which has different energy characteristics not captured by our model. 
When such architectural and energy efficiency differences are known, adding a user-supplied efficiency scaling factor to power predictions in \cref{eq:method-dynamic-power} could improve this exploration accuracy.

\subsection{Exploration: Algorithm Level} \label{sec:results-fusion}
\begin{figure}[t]
    \centering
    \includegraphics[width=\linewidth]{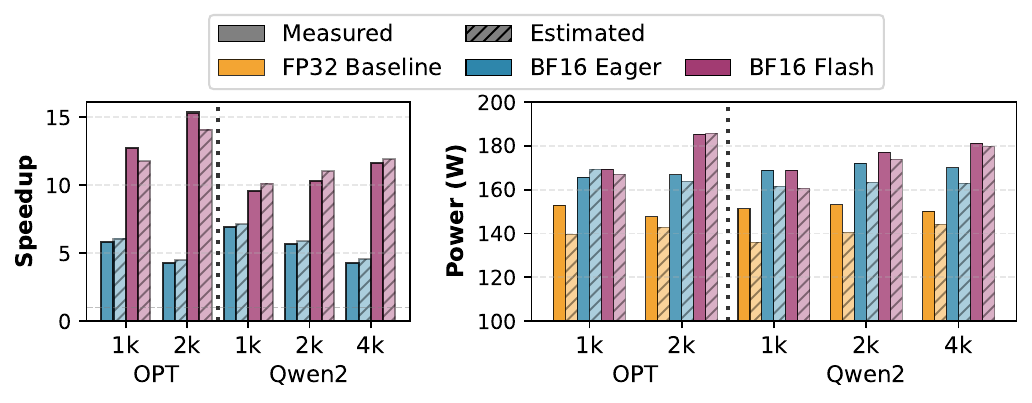}
    \caption{Estimation of (Left) the speedup achieved when changing bit precisions and using FlashAttention, and (Right) the power consumption for all three configurations.}
    \vspace{-2mm}
    \label{fig:results-fusion}
\end{figure}

AI model deployment requires choosing numerical precision (e.g., \texttt{fp32} vs. \texttt{bf16}) and backend kernels for self-attention (e.g., standard vs. FlashAttention), with potential trade-offs. 
We evaluate \name's ability to estimate these trade-offs for long-context language workloads with 1k+ sequence lengths on A100-40GB-PCIE. 
Using the \texttt{fp32} case as the baseline, we predict the speedup when: 1) the bit precision is changed to \texttt{bf16}, and 2) FlashAttention is enabled (\cref{fig:results-fusion}). 
\name accurately predicts the 4-5$\times$ speedup from \texttt{fp32}$\to$\texttt{bf16}, and the additional 2-3$\times$ speedup from FlashAttention.
Power estimation closely follows the ground truth across all three test cases, accurately capturing the increases in power consumption from the \texttt{fp32} baseline's 150W to 160-170W for \texttt{bf16} and above 180W for FlashAttention at long sequences. 
By quantifying these potential speedup-power trade-offs, our lightweight framework enables algorithm developers to gain hardware insights for these deployment decisions.

\subsection{Limitations}
\name assumes that kernels execute sequentially and exhibit regular execution patterns with predictable pipelining behavior. 
These assumptions enable fast predictions, but also limit its current applicability. 

First, workloads involving overlapping kernel executions, such as inter-GPU communication kernels, are not currently supported, as our model treats one kernel at a time.
Extending the framework to account for resource partitioning, such as SM and bandwidth allocation across concurrently executing kernels, can be a potential future direction. 
Second, kernels with irregular memory access patterns, such as those exploiting unstructured sparsity, challenge tile-based analysis, as tensors are stored in compressed formats rather than contiguous tiles. 
Statistical modeling approaches, which have shown promise for sparse accelerator performance modeling~\cite{sparseloop}, could be leveraged. 
Finally, \name currently targets single-GPU workloads. 
Extending predictions to multi-GPU settings with communication kernels modeled explicitly remains an important direction for future work.

\section{Related Work} \label{sec:related-work}

\noindent\textbf{GPU Power Models }
Several approaches with varying levels of architectural detail have been proposed. 
\cite{gpuwattch2013,accelwattch2021,ipp2010,gpujoule2019,guerreiro2018} provide power consumption breakdowns across architectural components and instruction types through micro-benchmarking. 
Black-box approaches \cite{wu2015,ali2023,zhang2024} abstract GPU internals and employ machine learning techniques for estimation. 
However, both approaches typically require hardware utilization information from either simulations or profiling (or runtime monitoring), which creates scalability issues.

\noindent\textbf{Analytical Models }
They predict hardware behavior from first principles.
Classic examples include the roofline model, which provides first-order performance~\cite{roofline} and energy~\cite{choi2013roofline} insights. 
An analytical approach has been explored for both general-purpose GPU modeling~\cite{hong2009} and workload-specific GPU modeling~\cite{Lym_2019,llmcompass}.
Also, analytical models for domain-specific accelerators have been proposed~\cite{timeloop,scalesim}, which are widely adopted for accelerator research. 



\noindent\textbf{Energy Optimization Techniques }
\cite{yu2023,dynamollm} showed that energy-aware resource allocation in datacenters, including GPU selection, DVFS, and parallelism strategies, improves energy efficiency of AI workloads. 
\cite{zeus2023,perseus2024} employed DVFS while minimizing performance slowdown by leveraging opportunities created by batched processing and pipeline parallelism in AI training. 
Additionally, \cite{energyawaretiling} investigated compiler-level optimizations that identify energy-optimal tiling strategies in GEMM workloads. 
While these solutions typically relied on preliminary energy profiling and runtime power monitoring, they could benefit from accurate power prediction provided by \name.

\section{Conclusion} \label{sec:conclusion}
This work presents \name, a fast and accurate GPU power estimation framework for AI workloads. 
\name tackles the core challenge of predicting the utilization information necessary for power modeling through a lightweight performance model, eliminating the scalability bottleneck of traditional power modeling approaches. 
It uses the coarse-grained execution patterns arising from software optimizations common in AI workloads to build an analytical scaffold, then applies data-driven refinement to achieve high accuracy.
This performance model construction allows utilization information to be naturally derived, enabling rapid power estimation. 
\name achieves 8\% average power estimation error across diverse AI workloads and strong exploration capabilities. 

\section*{Acknowledgments}
The authors thank Huamin Chen from RedHat for useful discussions.

\bibliographystyle{IEEEtran}
\bibliography{ref}

\appendix
\section{Artifact Appendix}

\subsection{Abstract}


We provide the artifacts of EnergAIzer, including 1) the source code of the estimation framework, 2) a pre-collected database for empirical fitting, and 3) ground-truth measurements to validate the predictions. 
Our artifacts include scripts that perform experiments needed to reproduce single kernel-level power and latency estimations (Fig. 7 and 10(a)) and end-to-end estimations for AI workloads (Fig. 9 and 11).
Beyond reproducing these key results, our artifacts can be adapted to make GPU power/energy predictions for diverse AI workloads and design space exploration of GPU configurations.
Running our artifacts requires a Python environment (i.e., Python3, virtual environment) in either x86-64 or Arm machines, and 200 MB of disk space. 
Unless the user intends to customize the artifacts for their own GPU database collection, GPU machines are not required. 

\subsection{Artifact check-list (meta-information)}


{\small
\begin{itemize}
  \item {\bf Algorithm}: Estimating GPU power and energy consumption for AI workloads using a lightweight approach for activity factor prediction.
  \item {\bf Program}: Python
  \item {\bf Run-time environment}: Miniconda Virtual Environment
  \item {\bf Hardware}: x86-64 or Arm machines (CPU)
  \item {\bf Experiments}: Kernel-level and end-to-end power and latency estimation for AI workloads.
  \item {\bf How much disk space required (approximately)?}: 200 MB
  \item {\bf How much time is needed to prepare workflow (approximately)?}: 10 minutes following the quick start script
  \item {\bf How much time is needed to complete experiments (approximately)?}: 2 hours
  \item {\bf Publicly available?}: Yes, available at {https://github.com/kyungmi-lee/energaizer-ispass26-artifact}
  \item {\bf Code licenses (if publicly available)?}: MIT
  \item {\bf Archived (provide DOI)?}: 10.5281/zenodo.18916559
\end{itemize}
}

\subsection{Description}

\subsubsection{How to access}

Please download the artifacts from the Github: {https://github.com/kyungmi-lee/energaizer-ispass26-artifact} or the Zenodo archive: {10.5281/zenodo.18916559}. 

\subsubsection{Software dependencies}

The provided bash scripts can be executed in Linux or Mac OS environments. 
The artifacts require Python3 and Anaconda/Miniconda Virtual Environments. 

\subsection{Installation}

The installation has two steps: 
1) download the pre-collected database for reproducing the results, and
2) building a virtual environment with dependent libraries.
Both steps can be performed with the provided scripts explained in \texttt{README.md}. 
Installation should take less than 10 minutes. 

\subsection{Experiment workflow}

The artifacts provide the scripts for running single kernel-level and end-to-end predictions. 
To run all experiments, simply run \texttt{bash figures/scripts/all.sh}. 
This script will execute three sub-experiments: 1) single kernel-level power and latency predictions, 2) end-to-end predictions for language and vision workloads, and 3) voltage-frequency scaling predictions for a few selected language workloads. 
Once this script is completed, which can take about 1-2 hours, open the Jupyter notebook \texttt{figures/figures.ipynb} and execute cells to plot the graphs.

\subsection{Evaluation and expected results}

The Jupyter notebook will plot these 4 figures:
\begin{itemize}
    \item Fig. 7: Validating kernel-level power predictions (GEMM, Softmax, FlashAttention) for NVIDIA A100 GPU at 900 MHz. 
    \begin{itemize}
        \item Expected results: For the three kernel types, the predicted power should closely follow the ground truth (i.e., the diagonal line in each graph). The mean average percentage error (MAPE) should be between 3-5\%. 
    \end{itemize}
    \item Fig. 9: Validating end-to-end latency and power predictions for language and vision workloads. We experiment with two GPUs, NVIDIA A100 and A10, both at 900 MHz. Also, we provide latency prediction comparisons with the state-of-the-art performance models for AI workloads. 
    \begin{itemize}
        \item Expected results: The average error rates of our latency and power predictions should be between 8-12\% for both GPUs. The compared prior models exhibit similar or higher error rates (latency only), between 9-30\%. Overall, the bar graph for each workload should closely resemble the figure in our paper. 
    \end{itemize}
    \item Fig. 10(a): Validating kernel-level latency predictions for NVIDIA A100 and A10, along with comparisons. 
    \begin{itemize}
        \item Expected results: Across all kernel types \textit{except Cnv} (TC - Tensor Core GEMM; CUDA - CUDA Core GEMM; Cnv - Convolution; Soft - Softmax; LN - Layer Normalization; EW - Elementwise), ours should exhibit $<10\%$ MAPE. For Conv, the error rate is around 20\%. The error rates of the compared prior models are expected to be higher for TC and CUDA GEMM kernels, at around 20\%. 
    \end{itemize}
    \item Fig. 11: Validating power predictions under voltage-frequency scaling for NVIDIA A100 GPU. 
    \begin{itemize}
        \item Expected results: For all 6 workloads, MAPEs should be between 6-13\%. In the graph, the estimated power should closely follow the ground truth. 
    \end{itemize}
\end{itemize}

\noindent For all experiments, we provide the ground truth measurements of GPU latency and power in \texttt{test/data/measurements}, allowing convenient artifact evaluation without requiring GPU access. 
However, users with access to an NVIDIA A100-40GB-PCIE GPU can run the measurements using the code and scripts provided in \texttt{test/code} as well. 

\subsection{Experiment customization}

We outline how to generally use our artifacts for future research in \texttt{README.md}. 
First, users can make predictions for their own AI workloads, following the input format EnergAIzer requires. 
The measurement code in \texttt{test/code} includes scripts to parse PyTorch/HuggingFace model descriptions into our input format. 
Second, users can evaluate voltage-frequency scaling behaviors for the range of frequencies. 
Finally, users can explore different GPU architecture configurations, such as forecasting the power consumption in other GPU generations (e.g., NVIDIA Hopper). 

\subsection{Notes}

We plan to extend our artifact repository to provide database collection scripts and support inter-GPU communication kernels.

\subsection{Methodology}

Submission, reviewing and badging methodology:

\begin{itemize}
  \item \url{https://www.acm.org/publications/policies/artifact-review-and-badging-current}
  \item \url{https://cTuning.org/ae}
\end{itemize}

\end{document}